\documentclass[a4paper]{article}

\usepackage[english]{babel}
\usepackage[utf8x]{inputenc}
\usepackage[T1]{fontenc}

\usepackage[a4paper,top=3cm,bottom=2cm,left=3cm,right=3cm,marginparwidth=1.75cm]{geometry}

\usepackage{bbm}
\usepackage{wrapfig}
\usepackage{listings}
\usepackage{color}
\definecolor{mygreen}{RGB}{28,172,0}
\definecolor{mylilas}{RGB}{170,55,241}
\usepackage{graphicx}
\usepackage{amsfonts}
\usepackage{booktabs}
\usepackage{siunitx}
\usepackage{caption}

\usepackage{float}
\usepackage{blindtext}
\usepackage{amsmath}
\usepackage{subcaption}
\usepackage{graphicx}
\usepackage{wasysym}
\usepackage{url}
\usepackage[colorlinks=true, allcolors=blue]{hyperref}

\numberwithin{equation}{subsection}
\newcommand{\RM}[1]{\MakeUppercase{\romannumeral #1{}}}
\newcommand{\ist}[1]{\overset{\footnotesize(\ref{#1})}{=}}
\newcommand{\iist}[2]{\overset{^{(\ref{#1})}}{\underset{^{(\ref{#2})}}{=}}}

\title{\textbf{About the solution of the numerical instability for topological solitons with long range interaction}}
\author{\textbf{Fabian Anmasser, Dominik Theuerkauf and Manfried Faber}}

\begin{document}
\maketitle

\begin{abstract}
The computations of solutions of the field equations in the Model of Topological Particles, formulated with a scalar SU(2)-field, have shown instabilities leading to discrepancies between the numerical and analytical solutions. We identify the origin of these deviations in misalignments of the rotational axes corresponding to the SU(2) elements. The system of a single soliton we use as an example to show that a constraint suppressing the wave-like disturbances is able to lead to excellent agreement between the result of the numerical minimisation procedure and the analytical solution.
\end{abstract}

\section{Introduction}

Topological solitons are interesting objects. Their masses are given by integrals over the energy density, the particle number is a topological quantum number and their interaction is a consequence of the topology. Well-known models of this type are the Sine-Gordon model and the Skyrme model. Both obey the laws of special relativity. The Sine-Gordon model~\cite{remoissenet:1999wa} is a 1+1D model with 1 degree of freedom and two types of solitons differing by their chirality acting like a charge. They behave as expected for extended charged particles. Particles of equal charge repel and of opposite charge attract each other and annihilate. Bound states of particle pairs oscillate and are therefore dubbed breathing modes.

Skyrme's model~\cite{Skyrme:1958vn,Skyrme:1961vq,mrs93} is formulated in 3+1D with the three degrees of freedom of an SU(2) field with the interpretation of the meson field in nuclei. It was intended as a model for particles with the strong, short range interaction of nucleons. Skyrme's model can not model the Coulomb interaction by topological properties.

In \cite{Faber:1999ia} a model in 3+1D with the three degrees of freedom of an SO(3) field was suggested, modelling a long range Coulomb interaction for topologically stable solitons. It is of similar spirit as the Skyrme model, but due to the different Lagrangian it allows for three topological quantum numbers $\pi_2(S^2)$,$\pi_3(S^2)$ and $\pi_3(S^3)$. The solitons of this model can be interpreted as Dirac monopoles~\cite{dirac:1931kp,dirac:1948um} without any singularity, without Dirac string and with a soft core. There are four types of stable solitons differing in the two topological quantum numbers $\pi_3(S^3)$ and $\pi_2(S^2)$ which can be interpreted as spin up and down, positive and negative charge. The equations of motion can be solved analytically for the one soliton systems. In addition, the model describes two types of Goldstone bosons, i.e. massless excitations propagating with the speed of light~\cite{Faber:2002nw}. In Ref.~\cite{Faber:1999ia} the model was published under the title ``model of topological fermions'' but possibly it should rather be dubbed \emph{model of topological particles} since it turned out that also the Goldstone bosons of the model are characterised by a topological quantum number~\cite{Faber:2017uvr}. Due to the non-linearity of the model more complicated systems have to be solved numerically. But the numerics has suffered from numerical instabilities leading to large uncertainties \cite{Wabnig,Resch,Theuerkauf}. In this article we report about a successful method to avoid these instabilities~\cite{Anmasser}. For the one soliton system we are able to compare in a careful analysis the numerical outcome with the exact analytical results. In this article we discuss the sources and the size of the errors of the numerical evaluations. In Sect.~\ref{Sec:Formulierung} we give a short overview of the model and the analytical solution for the one soliton system in Sect.~\ref{Sec:Soliton}. After discussing the cylindrical and the lattice formulation in Sects.~\ref{Sec:Zylinder} and \ref{Sec:Gitter} we demonstrate the failure of the calculation in Sect.~\ref{Sec:Genauigkeit}. Further, we suggest in this section, to improve the calculations by a constraint. With the improved numerics we then get good agreement with the analytical results and give characteristic numbers for the achieved accuracy.

\section{Formulation of the model}\label{Sec:Formulierung}

SU(2) is the double covering  group of SO(3). With the Rodriguez formula, SU(2) matrices can be expanded in a $\sin$ and a $\cos$ term, whereas the real $3\times 3$ SO(3) rotational matrices in 3D need three terms. It is therefore simpler to do the calculations in SU(2) than in SO(3). For smooth field configurations in space-time there is the only difference that for every configuration of the SO(3)-field of the model we get two SU(2) configurations, differing by a multiplication with the non-trivial centre element of SU(2). Taking this into account, the three degrees of freedom of the model are formulated with a scalar field of SU(2) matrices
\begin{equation}\label{unitquaternions}
Q(x)=q_0-\mathrm i\vec q(x)\vec\sigma\quad\textrm{with}\quad q_0^2+\vec q^2=1,
\end{equation}
of unit quaternions, where $\vec q\vec\sigma=q_i\sigma_i$ is an element of the su(2) algebra, with the usual Pauli matrices $\sigma_i$ and Einstein's summation convention applied.

The Lagrangian of the model reads~\cite{Faber:1999ia}
\begin{equation}
\mathcal{L}_{\mathrm{MTP}}=-\frac{\alpha_f\hbar c}{4\pi}
\left(\frac{1}{4}\vec{R}_{\mu\nu}\vec{R}^{\mu\nu}+\frac{q_0^{2m}}{r_0^4}\right),
\end{equation}
where $\alpha_f$ and $r_0$ are in principle arbitrary constants. Choosing for $\alpha_f$ the value of Sommerfeld's fine-structure constant the force field of solitons can be compared with the Coulomb field and the size parameter $r_0$ can be adjusted to the mass of the lightest fundamental, charged particles existing in nature, to electrons. The dynamical term proportional to $\vec{R}_{\mu\nu}\vec{R}^{\mu\nu}$ is equivalent to the Skyrme term of the Skyrme model. It can be formulated with the connection field on the SU(2) manifold $\vec\Gamma_\mu$ defined by
\begin{equation}\label{derivativeQ}
\partial_\mu Q=-\mathrm i\vec\Gamma_\mu\vec\sigma Q,\quad
\vec{\Gamma}_{\mu}(r,\varphi,z)\ist{unitquaternions}q_0\partial_\mu\vec q
-\vec q\partial_\mu q_0+\vec q\times\partial_\mu\vec q,
\end{equation}
$\vec R_{\mu\nu}$ is an area density on the SU(2) manifold
\begin{equation}\label{RSU2}
\vec R_{\mu\nu}=\vec\Gamma_\mu\times\vec\Gamma_\nu.
\end{equation}
This is valid in the gauge where the local coordinate systems on the SU(2) manifold are chosen as $\sigma_iQ$, see Eq.~(\ref{derivativeQ}), and where the Maurer-Cartan equation
\begin{equation}\label{MaurerCartan}
\partial_\mu\vec\Gamma_\nu-\partial_\nu\vec\Gamma_\mu=2\vec\Gamma_\mu\times\vec\Gamma_\nu,
\end{equation}
is satisfied. After a rotation of these local coordinates systems on the SU(2) manifold one recognizes in $\vec R_{\mu\nu}$ the well-known form of the field strength tensor in QCD and the curvature tensor in general relativity
\begin{equation}\label{RSU2gen}
\vec R_{\mu\nu}=\partial_\mu\vec\Gamma_\nu-\partial_\nu\vec\Gamma_\mu
-\vec\Gamma_\mu\times\vec\Gamma_\nu.
\end{equation}
The potential term proportional to $q_0^{2m}$ defines a two-fold degenerate vacuum at $q_0=0$. It fixes the size and mass of solitons.

We can relate the geometry to physics introducing a dual vector potential $\vec C_\mu$ and a dual field strength tensor $^*\vec F_{\mu\nu}$ by
\begin{equation}\label{sphericalVecPots}
\vec C_\mu=-\frac{e_0}{4\pi\varepsilon_0c}\vec\Gamma_\mu,\quad
^*\vec F_{\mu\nu}=-\frac{e_0}{4\pi\varepsilon_0c}\vec R_{\mu\nu}.
\end{equation}

Since we are considering static cases, we have no magnetic fields, $\vec B_i=0$ and get for the energy density, the 00 component of the energy-momentum tensor
\begin{equation}\label{EneDens}
\mathcal H=\underbrace{\frac{\epsilon_0}{2}\vec E_i\vec E_i}
_{\mathcal{H}_{\mathrm{cur}}}+\underbrace{\frac{\alpha_f\hbar c}{4\pi} \frac{q_0^{2m}}{r_0^4}}_{\mathcal{H}_{\mathrm{pot}}}.
\end{equation}

\subsubsection*{Scale dependencies}\label{HobDerr}
For stable, time independent solutions we get a condition for the stability of the solutions. The two terms in Eq.~(\ref{EneDens}) have different scale dependencies, $x\to\lambda x$
\begin{align}\label{KrummBeitr}
&H_\mathrm{cur}:=\int\mathrm d^3x\mathcal H_\mathrm{cur}\ist{EneDens}
\frac{\epsilon_0}{2}\int\mathrm d^3x\vec E_i\vec E_i\hspace{4mm}
\quad\rightarrow\quad
\frac{1}{\lambda}H_\mathrm{cur},\\
&H_\mathrm{pot}:=\int\mathrm d^3x\mathcal H_\mathrm{pot}
\ist{EneDens}\frac{\alpha_f\hbar c}{4\pi} \int\mathrm d^3x\frac{q_0^{2m}}{r_0^4}
\quad\rightarrow\quad\lambda^3H_\mathrm{pot}.\label{PotBeitr}
\end{align}
The stability of the total energy

\begin{equation}\label{GesamtE}
H_\mathrm{tot}:=H_\mathrm{cur}+H_\mathrm{pot}
\end{equation}
infers
\begin{equation}\label{EnergieBez}
\frac{\mathrm d}{\mathrm d\lambda}H_\mathrm{tot}\Bigr\rvert_{\lambda=1}
\ist{GesamtE}\frac{\mathrm d}{\mathrm d\lambda}H_\mathrm{cur}\Bigr\rvert
_{\lambda=1}+\frac{\mathrm d}{\mathrm d\lambda}H_\mathrm{pot}\Bigr\rvert_{\lambda=1}
\iist{KrummBeitr}{PotBeitr}-H_\mathrm{cur}+3H_\mathrm{pot}=0.
\end{equation}
This application of the Hobart-Derrick theorem~\cite{Hobart:1963rh,Derrick:1964gh} gives a useful check of the accuracy of numerical calculations. In Sect.~\ref{Sec:Genauigkeit} we will compare the numerical results with
\begin{equation}\label{Verh4}
\frac{H_\mathrm{tot}}{H_\mathrm{pot}}\iist{GesamtE}{EnergieBez}4.
\end{equation}

\section{Solitonic solution}\label{Sec:Soliton}

\begin{figure}[h!]
\centering
\begingroup
  \makeatletter
  \providecommand\color[2][]{%
    \GenericError{(gnuplot) \space\space\space\@spaces}{%
      Package color not loaded in conjunction with
      terminal option `colourtext'%
    }{See the gnuplot documentation for explanation.%
    }{Either use 'blacktext' in gnuplot or load the package
      color.sty in LaTeX.}%
    \renewcommand\color[2][]{}%
  }%
  \providecommand\includegraphics[2][]{%
    \GenericError{(gnuplot) \space\space\space\@spaces}{%
      Package graphicx or graphics not loaded%
    }{See the gnuplot documentation for explanation.%
    }{The gnuplot epslatex terminal needs graphicx.sty or graphics.sty.}%
    \renewcommand\includegraphics[2][]{}%
  }%
  \providecommand\rotatebox[2]{#2}%
  \@ifundefined{ifGPcolor}{%
    \newif\ifGPcolor
    \GPcolortrue
  }{}%
  \@ifundefined{ifGPblacktext}{%
    \newif\ifGPblacktext
    \GPblacktexttrue
  }{}%
  % define a \g@addto@macro without @ in the name:
  \let\gplgaddtomacro\g@addto@macro
  % define empty templates for all commands taking text:
  \gdef\gplbacktext{}%
  \gdef\gplfronttext{}%
  \makeatother
  \ifGPblacktext
    % no textcolor at all
    \def\colorrgb#1{}%
    \def\colorgray#1{}%
  \else
    % gray or color?
    \ifGPcolor
      \def\colorrgb#1{\color[rgb]{#1}}%
      \def\colorgray#1{\color[gray]{#1}}%
      \expandafter\def\csname LTw\endcsname{\color{white}}%
      \expandafter\def\csname LTb\endcsname{\color{black}}%
      \expandafter\def\csname LTa\endcsname{\color{black}}%
      \expandafter\def\csname LT0\endcsname{\color[rgb]{1,0,0}}%
      \expandafter\def\csname LT1\endcsname{\color[rgb]{0,1,0}}%
      \expandafter\def\csname LT2\endcsname{\color[rgb]{0,0,1}}%
      \expandafter\def\csname LT3\endcsname{\color[rgb]{1,0,1}}%
      \expandafter\def\csname LT4\endcsname{\color[rgb]{0,1,1}}%
      \expandafter\def\csname LT5\endcsname{\color[rgb]{1,1,0}}%
      \expandafter\def\csname LT6\endcsname{\color[rgb]{0,0,0}}%
      \expandafter\def\csname LT7\endcsname{\color[rgb]{1,0.3,0}}%
      \expandafter\def\csname LT8\endcsname{\color[rgb]{0.5,0.5,0.5}}%
    \else
      % gray
      \def\colorrgb#1{\color{black}}%
      \def\colorgray#1{\color[gray]{#1}}%
      \expandafter\def\csname LTw\endcsname{\color{white}}%
      \expandafter\def\csname LTb\endcsname{\color{black}}%
      \expandafter\def\csname LTa\endcsname{\color{black}}%
      \expandafter\def\csname LT0\endcsname{\color{black}}%
      \expandafter\def\csname LT1\endcsname{\color{black}}%
      \expandafter\def\csname LT2\endcsname{\color{black}}%
      \expandafter\def\csname LT3\endcsname{\color{black}}%
      \expandafter\def\csname LT4\endcsname{\color{black}}%
      \expandafter\def\csname LT5\endcsname{\color{black}}%
      \expandafter\def\csname LT6\endcsname{\color{black}}%
      \expandafter\def\csname LT7\endcsname{\color{black}}%
      \expandafter\def\csname LT8\endcsname{\color{black}}%
    \fi
  \fi
    \setlength{\unitlength}{0.0500bp}%
    \ifx\gptboxheight\undefined%
      \newlength{\gptboxheight}%
      \newlength{\gptboxwidth}%
      \newsavebox{\gptboxtext}%
    \fi%
    \setlength{\fboxrule}{0.5pt}%
    \setlength{\fboxsep}{1pt}%
\begin{picture}(5760.00,4320.00)%
    \gplgaddtomacro\gplbacktext{%
      \csname LTb\endcsname%
      \put(594,704){\makebox(0,0)[r]{\strut{}$0$}}%
      \put(594,2182){\makebox(0,0)[r]{\strut{}$0.1$}}%
      \put(594,3659){\makebox(0,0)[r]{\strut{}$0.2$}}%
      \put(726,484){\makebox(0,0){\strut{}$0$}}%
      \put(1499,484){\makebox(0,0){\strut{}$1$}}%
      \put(2272,484){\makebox(0,0){\strut{}$2$}}%
      \put(3045,484){\makebox(0,0){\strut{}$3$}}%
      \put(3817,484){\makebox(0,0){\strut{}$4$}}%
      \put(4590,484){\makebox(0,0){\strut{}$5$}}%
      \put(5363,484){\makebox(0,0){\strut{}$6$}}%
    }%
    \gplgaddtomacro\gplfronttext{%
      \csname LTb\endcsname%
      \put(3044,154){\makebox(0,0){\strut{}$r/r_0$}}%
      \put(3044,3989){\makebox(0,0){\strut{}radial energy densities}}%
      \csname LTb\endcsname%
      \put(4376,3486){\makebox(0,0)[r]{\strut{}radial contribution}}%
      \csname LTb\endcsname%
      \put(4376,3266){\makebox(0,0)[r]{\strut{}tangential contribtion}}%
      \csname LTb\endcsname%
      \put(4376,3046){\makebox(0,0)[r]{\strut{}potential contribution}}%
    }%
    \gplbacktext
    \put(0,0){\includegraphics{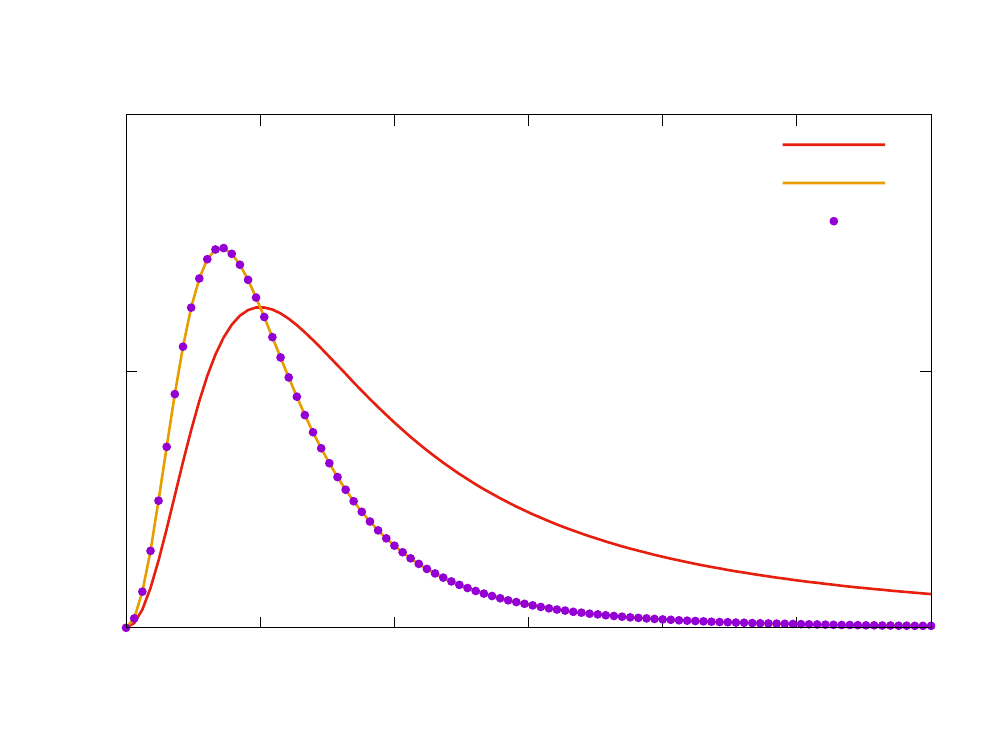}}%
    \gplfronttext
  \end{picture}%
\endgroup
\caption{Three contributions to the energy densities for the soliton solution with $m=3$. The radial contribution describes the Coulomb field, regularised by the structure of the model.}
\label{enedens3}
\end{figure}

For static monopoles at the origin~\cite{Faber:1999ia} the model reduces to spherical symmetry in 3D with spherical coordinates $r,\vartheta,\varphi$~\footnote{We would like to emphasize that we use the arrow symbol like $\vec n$ for vectors in the su(2) algebra and bold symbols like $\mathbf r$ for vectors in space.}
\begin{equation}\label{RegularIgel}
q_0=\cos\alpha(r),\quad\vec q=\vec n({\mathbf r})\sin\alpha(r),\quad
n_i({\mathbf r})=\frac{x_i}{r},\quad\alpha(r)\in[0,\frac{\pi}{2}],
\end{equation}
where $\mathbf r$ denotes a vector in space and $\vec n$ a unit vector in the su(2) algebra. The Euler-Lagrange equation, a non-linear differential equation
\begin{equation}\label{nlDE}
\partial^2_\rho\cos\alpha+\frac{(1-\cos^2\alpha)\cos\alpha}{\rho^2}
 -m\rho^2\cos^{2m-1}\alpha=0\quad\textrm{with}\quad\rho:=\frac{r}{r_0}
\end{equation}
has for $m=3$ a simple solution
\begin{equation}\label{MinLoesm3}
\tan\alpha(r)=\rho.
\end{equation}
Due to the simplicity of this solution we will use further on the case $m=3$. 
Corresponding to the three terms in Eq.~(\ref{nlDE}) the radial energy density has three contributions
\begin{equation}\label{radialdensities}
h(\rho)=\frac{\alpha_f\hbar c}{r_0}\Big[\frac{\rho^2}{2(1+\rho^2)^2}
 +\frac{\rho^2}{(1+\rho^2)^3}+\frac{\rho^2}{(1+\rho^2)^3}\Big],
\end{equation}
a radial field from the contributions of $\vec R_{\vartheta\varphi}$ approaching at large distances the Coulomb field of a point charge, a tangential field from $\vec R_{r\vartheta}$ and $\vec R_{r\varphi}$ and a potential contribution. They are depicted in Fig.~\ref{enedens3}. Integrating this radial energy density over $\rho$ we get the energy for the monopole. Comparing this energy to the rest energy of the lightest fundamental monopole existing in nature~\footnote{The most accurate value known up to now is $m_ec^2=0.510 998 950 00(15)$ MeV. Hence 0.511 MeV is a very good approximation to the experimental value.} we can fix the radius $r_0$, defining the size of the soliton~\cite{Faber2017}
\begin{equation}
\label{eqn:analyLsg}
H_\mathrm{mono}=\int_0^\infty\mathrm d\rho\,h(\rho)\ist{radialdensities}
\frac{\alpha_f\hbar c}{r_0}\frac{\pi}{4}=0.511\,\mathrm{MeV}\quad
\rightarrow r_0=2.21\,\mathrm{fm}.
\end{equation}

\subsubsection*{Electrodynamic limit}\label{sec:EnergOutside}

As long as we do not have analytical solutions for static systems with several charges or scattering problems, we are able to treat the fields outside the numerical integration region within classical electrodynamics only, a scenario which we dubbed in Ref.~\cite{Faber:2002nw} electrodynamic limit.

In this limit we assume $q_0=0$ and neglect therefore tangential and potential energy contributions. To get estimates of the accuracy of the numerical calculations, it is sufficient to determine the size of these neglected contributions in spherical coordinates in the region $\rho=\frac{r}{r_0}>\rho_>=\frac{r_>}{r_0}$. Their contributions to the radial energy density are given by the integrals over the second and third term in Eq.~(\ref{radialdensities}), by

\begin{equation}\begin{aligned}\label{errortp}
\frac{\alpha_f\hbar c}{r_0}\int_{\rho_>}^\infty\frac{2\rho^2}{(1+\rho^2)^3}\;
\mathrm d\rho&\ist{radialdensities}\frac{\alpha_f\hbar c}{r_0}\frac{\pi}{8}
\left[1-\frac{2}{\pi}\left(\arctan\;\rho_>
+\frac{\rho_>(\rho_>^2-1)}{(\rho_>^2+1)^2}\right]\right)\longrightarrow\\
&\overset{\footnotesize \rho_>\to\infty}{\longrightarrow}
\frac{\alpha_f\hbar c}{r_0}\frac{\pi}{8}\cdot\frac{16}{3\pi}\,\rho_>^{-3}.
\end{aligned}\end{equation}
Dividing by the integral over the whole $r$-axis, $\frac{\alpha_f\hbar c}{r_0}\frac{\pi}{8}$, results in the relative error of the sum of these contributions
\begin{equation}\label{elrelerr}
\Delta_\mathrm{el}(\rho_>):=1-\frac{2}{\pi}\left(\arctan\;\rho_>
+\frac{\rho_>(\rho_>^2-1)}{(\rho_>^2+1)^2}\right)
\overset{\footnotesize \rho_>\to\infty}{\longrightarrow}
\frac{16}{3\pi}\,\rho_>^{-3}.
\end{equation}
\begin{figure}[h!]
\centering
\includegraphics[scale=0.5]{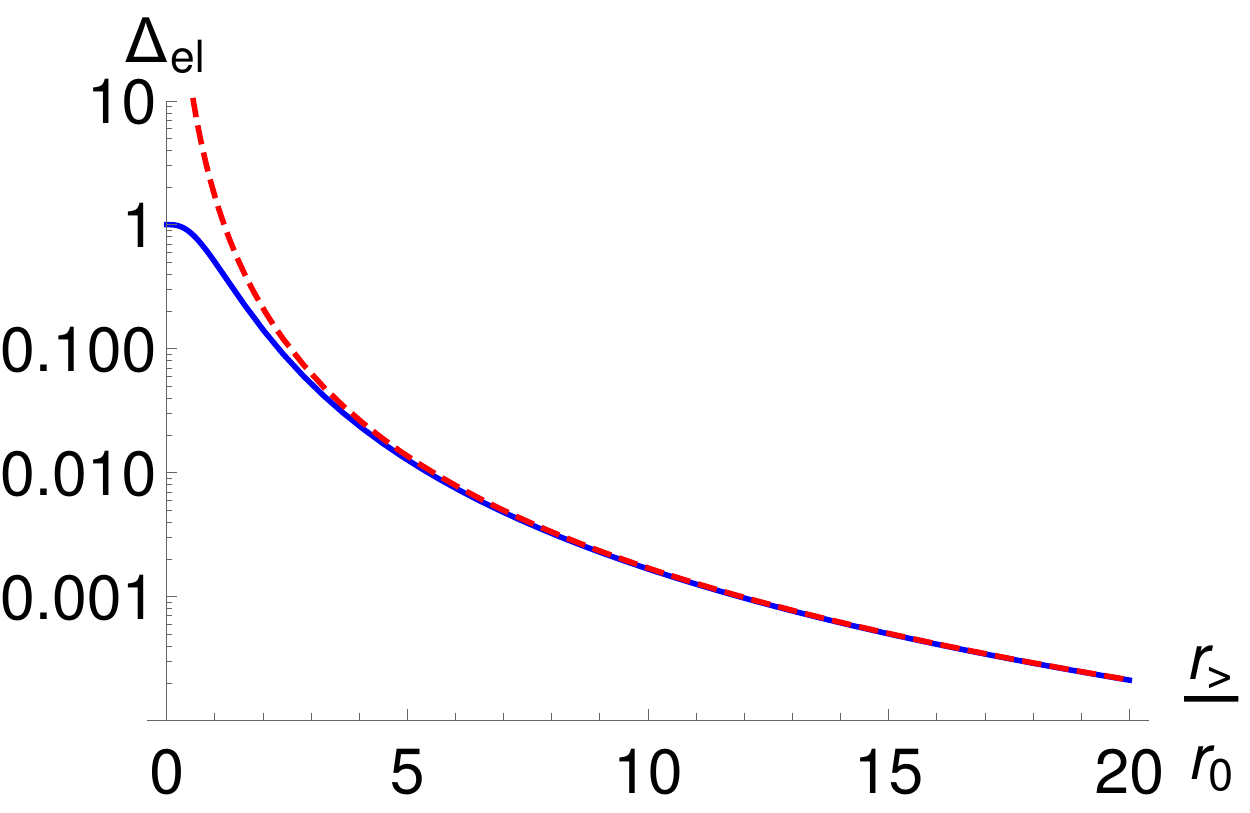}
\caption{Relative error $\Delta_\mathrm{el}$ in electrodynamic limit, when tangential and potential energy contributions are neglected (full blue line) for $r>r_>$. For large $r_>$ this relative error decays with the third power of $r_>/r_0$ (dashed red line), see Eq.~(\ref{elrelerr}).}
\label{elrelError}
\end{figure}
In the further calculations we are using $r_>=10~r_0$ with an error of 0.167\,\%. From Fig.~\ref{elrelError} we can see that for increasing $r_>$ the relative error decreases nicely with $(r_>/r_0)^{-3}$.

For comparison, in the same region the first term in Eq.~(\ref{radialdensities}) contributes with 12.6\,\%, which should not be neglected and is taken care of in the electrodynamic limit. These contributions we are going to determine in cylindrical coordinates.

\section{Cylindrical formulation}\label{Sec:Zylinder}
The algorithm presented here \cite{Wabnig,Theuerkauf,Anmasser}, uses cylindrical coordinates to work with, because it should be capable of computing dipoles as well. Therefore, we will also do monopole calculations in cylindrical coordinates $r,\varphi,z$ in order to get insight in the accuracy of the numerical calculations.
The general soliton field~(\ref{unitquaternions}) for any configuration in cylindrical coordinates reads
\begin{equation}\label{equ:SU2SolitonFeld}
Q(r,\varphi,z)\ist{unitquaternions}q_0(r,z)-i\vec\sigma\vec q(r,\varphi,z),\quad
\text{with}\quad\vec q(r,\varphi,z)=
\begin{pmatrix}q_r(r,z)\cos\varphi\\q_r(r,z)\sin\varphi\\q_z(r,z)
\end{pmatrix}.
\end{equation} 
For the affine connection~(\ref{derivativeQ}) we obtain
\begin{align}
\vec\Gamma_r&\ist{derivativeQ}\begin{pmatrix}
(q_0\partial_rq_r-q_r\partial_rq_0)\cos\varphi
+(q_r\partial_rq_z-q_z\partial_rq_r)\sin\varphi\\
(q_0\partial_rq_r-q_r\partial_rq_0)\sin\varphi
+(q_z\partial_rq_r-q_r\partial_rq_z)\cos\varphi\\
q_0\partial_rq_z- q_z\partial_rq_0\end{pmatrix}\\
\vec\Gamma_\varphi&\ist{derivativeQ}q_r
\begin{pmatrix}-q_0\sin\varphi-q_z\cos\varphi\\q_0\cos\varphi-q_z\sin\varphi\\
q_r\end{pmatrix}\\
\vec\Gamma_z&\ist{derivativeQ}
\begin{pmatrix}(q_r\partial_zq_z-q_z\partial_zq_r)\sin\varphi
+(q_0\partial_zq_r-q_r\partial_zq_0)\cos\varphi\\
(q_z\partial_zq_r-q_r\partial_zq_r)\cos\varphi
+(q_0\partial_zq_r-q_r\partial_zq_0)\sin\varphi\\
q_0\partial_zq_z-q_z\partial_zq_0\end{pmatrix}.
\end{align}
The components of the curvature tensor~(\ref{RSU2}) read
\begin{align}
\vec R_{\varphi z}&\ist{RSU2}q_r
\begin{pmatrix}\partial_zq_0\sin\varphi
+\partial_zq_z\cos\varphi\\-\partial_zq_0\cos\varphi
+\partial_zq_z\sin\varphi\\-\partial_zq_r\end{pmatrix},\\
\vec R_{zr}&\ist{RSU2}\frac{\partial_rq_r\partial_zq_z
-\partial_zq_r\partial_rq_z}{q_0}\begin{pmatrix}
-q_0\sin\varphi-q_z\cos\varphi\\q_0\cos\varphi-q_z\sin\varphi\\ q_r\end{pmatrix},\\
\vec R_{r \varphi}&\ist{RSU2}q_r
\begin{pmatrix}-\partial_rq_0\sin\varphi-\partial_rq_z\cos\varphi\\
\partial_rq_0\cos\varphi-\partial_rq_z\sin\varphi\\\partial_rq_r\end{pmatrix}.
\end{align}
and its squares
\begin{equation}\begin{aligned}\label{QuadR}
\vec R_{\varphi z}^2&
=q_r^2\big[(\partial_zq_0)^2+(\partial_zq_r)^2+(\partial_zq_z)^2\big],\\
\vec R_{zr}^2&=
q_0^{-2}\big(\partial_rq_r\partial_zq_z-\partial_zq_r\partial_rq_z\big)^2,\\
\vec R_{r \varphi}^2 &=
q_r^2\big[(\partial_rq_0)^2+(\partial_rq_r)^2+(\partial_rq_z)^2 \big].
\end{aligned}\end{equation}
With the proper conversion factor to SI units and the length scales for cylindrical coordinates, $l_{\varphi}=r \text{~and~} l_{r}=l_{z}=1$ we adjust relation~(\ref{sphericalVecPots}) between curvature tensor and electric field strength to cylindrical coordinates
\begin{equation}\label{Ezylinder}
\vec E_r=-\frac{e_0}{4\pi\epsilon_0}\frac{1}{l_\varphi l_z}\vec R_{\varphi z},\quad
\vec E_\varphi=\frac{e_0}{4\pi\epsilon_0}\frac{1}{l_rl_z}\vec R_{r z},\quad
\vec E_z=-\frac{e_0}{4\pi\epsilon_0}\frac{1}{l_rl_\varphi}\vec R_{r \varphi}.
\end{equation}
This leads to the density of the curvature energy
\begin{equation}\begin{split}\label{equ:curvatureEnergy}
\mathcal H_\mathrm{cur}\ist{EneDens}
\frac{\epsilon_0}{2}&\big(\vec E_r^2+\vec E_\varphi^2+\vec E_z^2\big)
\iist{QuadR}{Ezylinder}\frac{\alpha_f\hbar c}{8\pi}\frac{1}{r^2}
\bigg\{q_r^2\big[(\partial_zq_0)^2+(\partial_zq_r)^2++(\partial_zq_z)^2\big]\\
&+\frac{r^2}{q_0^2}\big(\partial_rq_r\partial_zq_z
-\partial_zq_r\partial_rq_z\big)^2+q_r^2\big[(\partial_rq_0)^2
+(\partial_rq_r)^2+(\partial_rq_z)^2\big]\bigg\},
\end{split}\end{equation}
where we used the definition of the fine structure constant $\alpha_f:=e_0^2/(4\pi\epsilon_0\hbar c)$.

\section{Lattice computation}\label{Sec:Gitter}

We consider the case of a monopole at the centre of a cylinder. Inside this cylinder we introduce the dimensionless coordinates $\bar{r}, \bar{\varphi}, \bar{z} \in \mathbb{Z}$, defined by the relations
\begin{equation}\label{eqn:DimensionlessCord}
r=a\bar r,\quad\varphi=\bar \varphi,\qquad z=a\bar z,\quad
\bar r\in\{0,1,\dots,n_r\},\quad \bar z\in\{-n_z,-n_z+1,\dots,\,n_z\}
\end{equation}
who just number the lattice points. Due to the rotational symmetry around the z-axis we perform the $\varphi$-integrations analytically and are left with a two dimensional lattice with spacing $a$, often referred to as box in the following. It is characterized with the number $n_r$ of points in $r$ direction and the number $n_z$ of points in $\pm z$ direction. Such a finite lattice suffers from boundary effects. In the example of lattice QCD the boundary problems are often diminished by periodic boundary conditions.  In the present model with solitons with long range Coulomb interaction periodic boundary conditions are in general not useful. Since we know the field configurations of charged particles from classical electrodynamics analytically, we use them for the boundary conditions.

\subsubsection{Outside electric energy}

\begin{wrapfigure}[12]{R}{5cm}
\vspace{-8mm}
\centering
\includegraphics[scale=0.4]{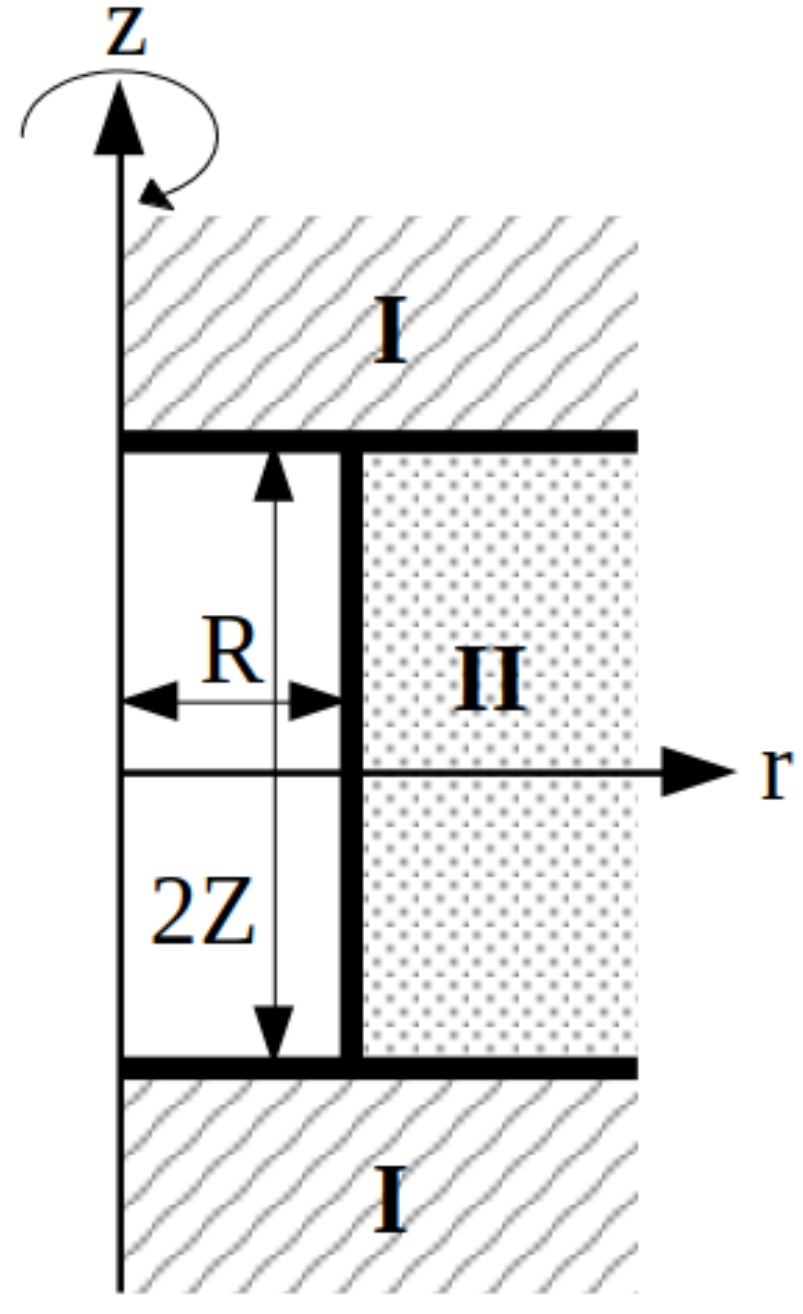}
\caption{Integration areas}
\label{fig:IntegrationAreas}
\end{wrapfigure}
For a monopole with unit charge $e_0$ at the centre of a cylinder the electric field strength reads within Maxwell's theory
\begin{equation}\label{EMaxwell}
\vec E(r)=\frac{e_0}{4\pi\epsilon_0}\frac{\vec r}{r^3}
=\frac{1}{4\pi\epsilon_0}\frac{e_0}{\sqrt{(r^2+z^2)^3}}
\begin{pmatrix}r\\0\\z\end{pmatrix}.
\end{equation}
We have to integrate $|\vec{E}|^2$ over the whole volume except the cylinder, characterised by its radius $R$ and its half length $Z$. Due to the cylindrical symmetry, we get an integration factor of $2\pi$ and we are able to restrict our consideration to the $rz$-plane, where the cylinder gets projected to a rectangular box. We split the remaining two dimensional area into three smaller regions, which are shown in figure \ref{fig:IntegrationAreas}. The energy outside the box may be written as
\begin{equation}\begin{aligned}\label{Helout}
H^\mathrm{out}_\mathrm{el}&\ist{EneDens}\frac{\epsilon_0}{2}
\int_{\mathbb R^3\setminus\text{box}}\mathrm d^3x|\vec{E}|^2
=\frac{\epsilon_0}{2}\int_{\mathbb R^3\setminus\text{box}}
r\,\mathrm dr\,\mathrm d\varphi\,\mathrm dz|\vec{E}|^2=\pi\epsilon_0
\int_{\mathbb R^2\setminus \text{box}}r\,\mathrm dr\,\mathrm dz|\vec E|^2=\\
&=\pi\epsilon_0\bigg\{\underbrace{\int_{-\infty}^{-Z}\int_0^\infty r\,
\mathrm dr\,\mathrm dz}_\textbf{\RM{1}}+\underbrace{\int_{-Z}^Z\int_R^\infty
r\,\mathrm dr\,\mathrm dz}_\textbf{\RM{2}}
+\underbrace{\int_Z^\infty\int_0^\infty
r\,\mathrm dr\,\mathrm dz}_\textbf{\RM{1}}\bigg\}|\vec{E}|^2=\\
&\ist{EMaxwell}\frac{\alpha_f\hbar c}{4}\bigg(\frac{1}{Z}
+\frac{1}{R}\arctan\frac{Z}{R}\bigg).
\end{aligned}\end{equation}

\subsubsection{Outside potential energy}

After we did the appraisal for the involved energies in- and outside a sphere in Sect.~\ref{sec:EnergOutside}, we want go on and find an analytical expression for the potential energy outside the cylindrical box, which we neglect in the electrodynamic limit. We integrate over $\mathcal{H}_{\mathrm{pot}}$ in Eq.~(\ref{EneDens})
\begin{align}
H_\mathrm{pot}^\mathrm{out}&\ist{EneDens}\frac{\alpha_f\hbar c}{4\pi r_0^4}
\int_{\mathbb R^3\setminus\text{box}}\mathrm d^3x\,q_0^6\iist{RegularIgel}{MinLoesm3}
\frac{\alpha_f\hbar c}{4\pi r_0^4} \int_{\mathbb{R}^3\setminus\text{box}}r\,\mathrm dr\,\mathrm d\varphi\,\mathrm{d}z\,\frac{1}{[1+(r^2+z^2)/r_0^2]^3}=\\
&=\frac{\alpha_f\hbar c}{2r_0}\frac{1}{4}\bigg\{
\arctan\frac{1}{Z_0}-\frac{Z_0}{1+Z_0^2}+\frac{Z_0}{(1+R_0^2)(1+R_0^2+Z_0^2)}
+\frac{1}{[1+R_0^2]^{3/2}}\arctan\frac{Z_0}{\sqrt{1+R_0^2}}\bigg\},\nonumber\\
&\hspace{20mm}\textrm{with}\quad Z_0=\frac{Z}{r_0},\;R_0=\frac{R}{r_0},\nonumber
\end{align}
where we were splitting up the integral in the exact same way as done in (\ref{Helout}) for the electrical energy outside the box.

If we consider a lattice with the specifications of $Z=10\,r_0$ and $R=10\,r_0$ we get a relative error of 0.122\,\%. This value is a little smaller than the error 0.167\,\%, which we got for a spherical volume in Eq.~(\ref{elrelerr}). The potential energy outside the box contributes therefore to the total energy with
\begin{equation}\label{schaetzwertPotOut}
H_{\mathrm{pot}}^{\mathrm{out}} \approx 0.156\,\mathrm{keV},
\end{equation}
which is about 0.31\,\permil\,of the rest mass of an electron, $m_ec^2\approx0.511\,\mathrm{MeV}$.

\subsection{Discretization}\label{sec:Discret}

Having gathered all energy contributions, we are now ready to go on with the discretization, which is necessary for the numerical calculations. The soliton field $Q$ stated in (\ref{equ:SU2SolitonFeld}), with its components $q_0$, $q_r$ and $q_z$ is only defined on the sites of the $rz$-lattice.

\subsubsection*{Derivatives}

To evaluate the curvature energy density $\mathcal{H}_{cur}$ (\ref{equ:curvatureEnergy}), we need the soliton field's derivatives of the form $\partial_i q_j$ in $r$ and $z$ directions only. Labelling the points in each of the two directions with integers as indicated in Eq.~(\ref{eqn:DimensionlessCord}) we are using the five-point method for the first derivative
\begin{equation}
\frac{\mathrm{d}f}{\mathrm{d}x_i}\approx\frac{f(x_{i-2})-8f(x_{i-1})+8f(x_{i+1})-f(x_{i+2})}{12a},
\end{equation}
whenever possible. It is correct up to $a^4$. For boundary points only left or right derivatives and at neighbouring points second order approximations are used.

With these expressions, we are prepared to calculate the curvature energy density $\mathcal{H}_{\mathrm{cur}}$ numerically.

\subsubsection*{Energies on the lattice}\label{sec:EnergiesOnTheLattice}

These derivatives are applied onto $\mathcal{H}_{\textrm{cur}}$ of Eq.~(\ref{equ:curvatureEnergy}) to get the curvature energy on the lattice. After performing the $\varphi$ integration analytically and using the dimensionless coordinates $\bar r$ and $\bar z$ of Eq.~(\ref{eqn:DimensionlessCord}) we can write
\begin{equation}\label{eqn:HcuramLattice}
H_\mathrm{cur}^\mathrm{box}\iist{equ:curvatureEnergy}{eqn:DimensionlessCord}
2\pi\,a^3
\int_\mathrm{box}\bar r\,\mathrm d\bar r\,\mathrm d\bar z\,\mathcal H_\mathrm{cur}
=:\frac{\alpha_f\hbar c}{a}\bar H_\mathrm{cur}^\mathrm{box}.
\end{equation}
From the potential energy density (\ref{EneDens}) we get similarly for the potential energy on the lattice
\begin{equation}\label{eqn:HpotamLattice}
H_\mathrm{pot}^\mathrm{box}\ist{EneDens}2\pi\frac{\alpha_f\hbar c}{4\pi\bar r_0^4}
\frac{1}{a}\int_\mathrm{box}\bar r\,\mathrm d\bar r\,\mathrm d\bar z\,q^{2m}_0
=:\frac{\alpha_f\hbar c}{a}\underbrace{\frac{1}{2\bar r_0^4}\int_\mathrm{box}
\bar r\,\mathrm d\bar r\,\mathrm d\bar z\,q_0^{2m}}_{\bar H_\mathrm{pot}^\mathrm{box}}
\end{equation}
with the soliton radius
\begin{equation}\label{solRadLat}
\bar r_0:=\frac{r_0}{a}
\end{equation}
in lattice units.
Therefore, the entire energy inside the box reads
\begin{equation}
\label{eqn:energyInbox}
H^\mathrm{box}\iist{eqn:HcuramLattice}{eqn:HpotamLattice}\frac{\alpha_f\hbar c}{a}\bigg(\bar H_\mathrm{cur}^\mathrm{box}+\bar H_\mathrm{pot}^\mathrm{box} \bigg).
\end{equation}
This energy is computed on the lattice and minimised by a conjugate gradient descent algorithm. Here, we want to emphasize again, that quantities with a bar over them are always dimensionless as can be seen in (\ref{eqn:energyInbox}), because $\alpha_f$ has no dimension, $\hbar c\approx 200\,\mathrm{MeV~fm}$ and $a$ measures the distance between two neighbouring points in fm. To enhance the accuracy of our calculations, we decided to do a cubic interpolation of the curvature- and potential energy density between the lattice points before integrating them.

Last, but not least we express the electrical energy outside the box $H^\mathrm{out}_\mathrm{el}$ in terms of dimensionless units
\begin{equation}\label{ElEneAus}
H_\mathrm{el}^\mathrm{out}\ist{Helout}\frac{\alpha_f\hbar c}{4}\bigg(\frac{1}{Z}
+\frac{1}{R}\arctan\frac{Z}{R}\bigg)
=\frac{\alpha_f\hbar c}{a}\underbrace{\frac{1}{4}\bigg(\frac{1}{\bar Z}
+\frac{1}{\bar R}\arctan\frac{\bar Z}{\bar R}\bigg)}_{\bar H^\mathrm{out}_\mathrm{el}}.
\end{equation}
Summing up the various energy contributions, the total energy $H_{\mathrm{tot}}$, which will be minimised, reads
\begin{equation}\label{GesamtEne}
H_\mathrm{tot}=H^\mathrm{box}+H^\mathrm{out}\iist{eqn:energyInbox}{ElEneAus}
\frac{\alpha_f\hbar c}{a}
\bigg(\bar H_\mathrm{cur}^\mathrm{box}+\bar H_\mathrm{pot}^\mathrm{box}
+\bar H_\mathrm{el}^\mathrm{out}\bigg)
=\frac{\alpha_f\hbar c}{a}\bar H_\mathrm{tot}.
\end{equation}

\section{Results and accuracy}\label{Sec:Genauigkeit}

The presented algorithm essentially consists of two different parts, both carrying a numerical error. The first part is to calculate the various energy components and the second one is the energy minimisation to find the associated configuration. We realized, that we have to be very careful to avoid numerical instabilities, therefore we will take a look at both steps separately in the following.

\subsection{Precision of the energy computation}\label{PrecisionOfTheEnergyComputations}

The computation of the energy for a monopole with radius $\bar r_0$ in lattice units $a$ results in $\bar H_\mathrm{tot}$, the total energy $H_\mathrm{tot}$ in units of $\frac{\alpha_f\hbar c}{a}$, see Eq.~(\ref{GesamtEne}). We can fix the lattice spacing $a$ from
\begin{equation}\label{GitKonst}
a\ist{solRadLat}\frac{r_0}{\bar r_0}\ist{eqn:analyLsg}
\frac{2.21\mathrm{fm}}{\bar{r}_0}.
\end{equation}
Now we are capable of transforming the total energy on the lattice $\bar H_\mathrm{tot}$ into common energy units
\begin{equation}
H_\mathrm{tot}\ist{GesamtEne}\frac{\alpha_f\hbar c}{a}\bar H_\mathrm{tot}
\ist{GitKonst}\frac{1.44}{2.21}\,\bar r_0\,\bar H_\mathrm{tot}\,\mathrm{MeV},
\end{equation}
which we are able to compare with the analytical value. Beside the total energy we compare the ratio $H_{\mathrm{tot}}/H_{\mathrm{pot}}$ with the analytical result~(\ref{Verh4}). With these two measures, we are now ready to do some simulations and check the results for their accuracy.\newline
\begin{figure}[h!]
\centering
\includegraphics[width=0.8\textwidth]{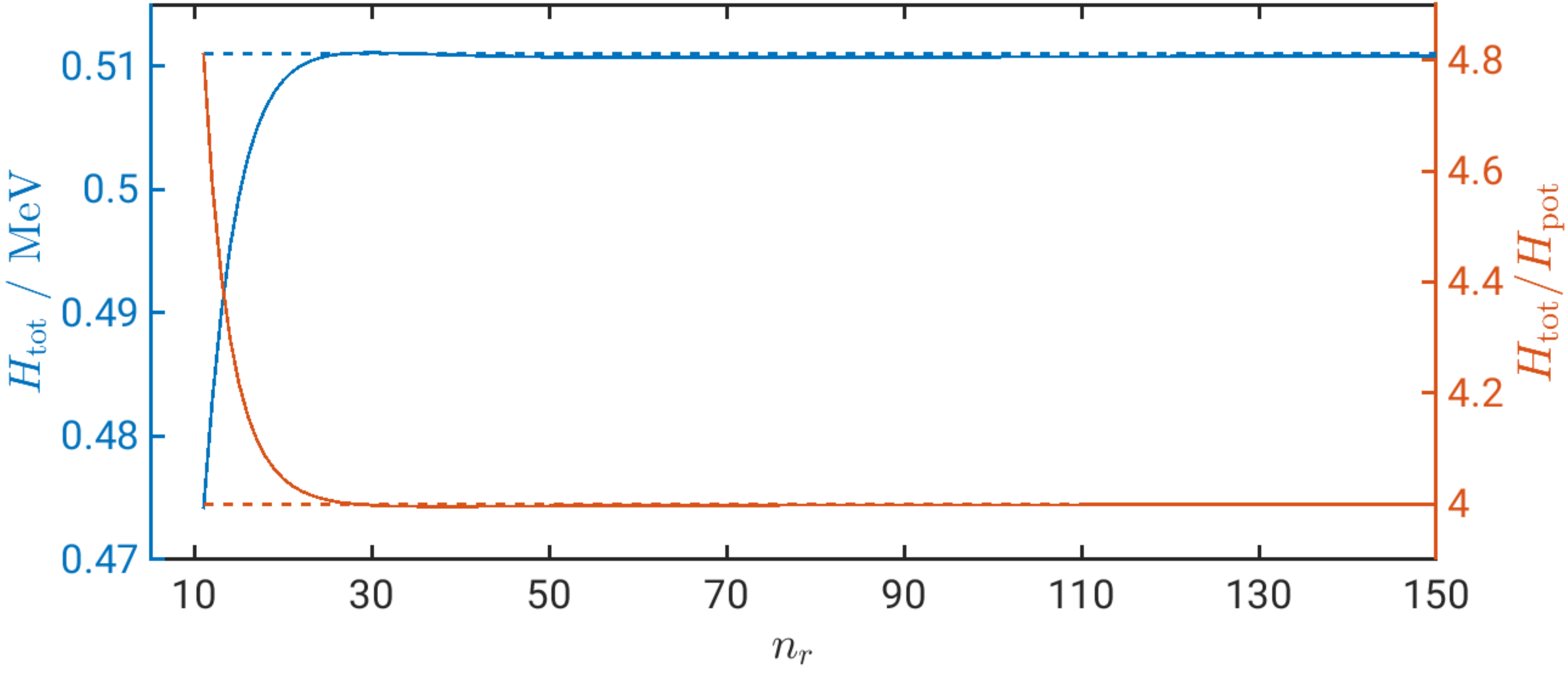}
\caption{Total energy and ratio $H_\mathrm{tot}/H_\mathrm{pot}$ for different lattices of size $n_r\times2n_z$ with $n_r=n_z$. Analytical values are marked with dashed lines. The plot shows the numerical results for a monopole configuration placed on the lattice. No minimisation at this point.}
\label{fig:EnergyCompPrecision}
\end{figure}
We placed the analytical solution of the monopole configuration on the lattice and calculated the energy contributions as described in section \ref{sec:EnergiesOnTheLattice}. The results before any minimisation are plotted in Fig.~\ref{fig:EnergyCompPrecision}, where the two measures of accuracy are plotted versus different grid sizes. Blue lines indicate the total energy $H_\mathrm{tot}$ and brown lines the ratio $H_\mathrm{tot}/H_\mathrm{pot}$. The analytical values are marked with dashed lines. As expected, for bigger lattices the calculations become more accurate. Below we will discuss the results after minimisation.

\subsection{Precision of the conjugate gradient minimisation}\label{PrecisionOfTheConjugateGradientMinimization}

Up to now, all we did was to place the centre of the analytical monopole configuration at the origin of the lattice and calculate its total energy. Carrying on, we want to test the minimising procedure\footnote{The minimisation algorithm is taken from \url{https://de.mathworks.com/matlabcentral/fileexchange/75546-conjugate-gradient-minimisation}. It is a more dimensional conjugate gradient method to find a local minimum of a function depending on many arguments, $f=f(x_1, x_2, ..., x_n)$.}. If there were no numerical errors, the algorithm would stop after a few iterations. Every deviation from the original configuration may be seen as an error.\newline
In our case, the function to be minimised is the total energy $\bar H_\mathrm{tot}$ which depends on the field components
\begin{equation}
  \bar H_\mathrm{tot}=\bar H_\mathrm{tot}\big(q_0(\bar r,\bar z),q_r(\bar r,\bar z),q_z(\bar r,\bar z)\big)\quad\textrm{with}\quad q_0^2+q_r^2+q_z^2=1.
\end{equation}
The conjugate gradient method starts at a given point on the hypersurface of $\bar{H}_{\mathrm{tot}}$, which in our case is associated with the analytical configuration and iterates its way to a local minimum. After the difference of energies of two consecutive iterations falls below a certain value, the algorithm stops.

It turns out that the procedure fails.

\begin{figure}[H]
\centering
\includegraphics[width=0.8\textwidth]{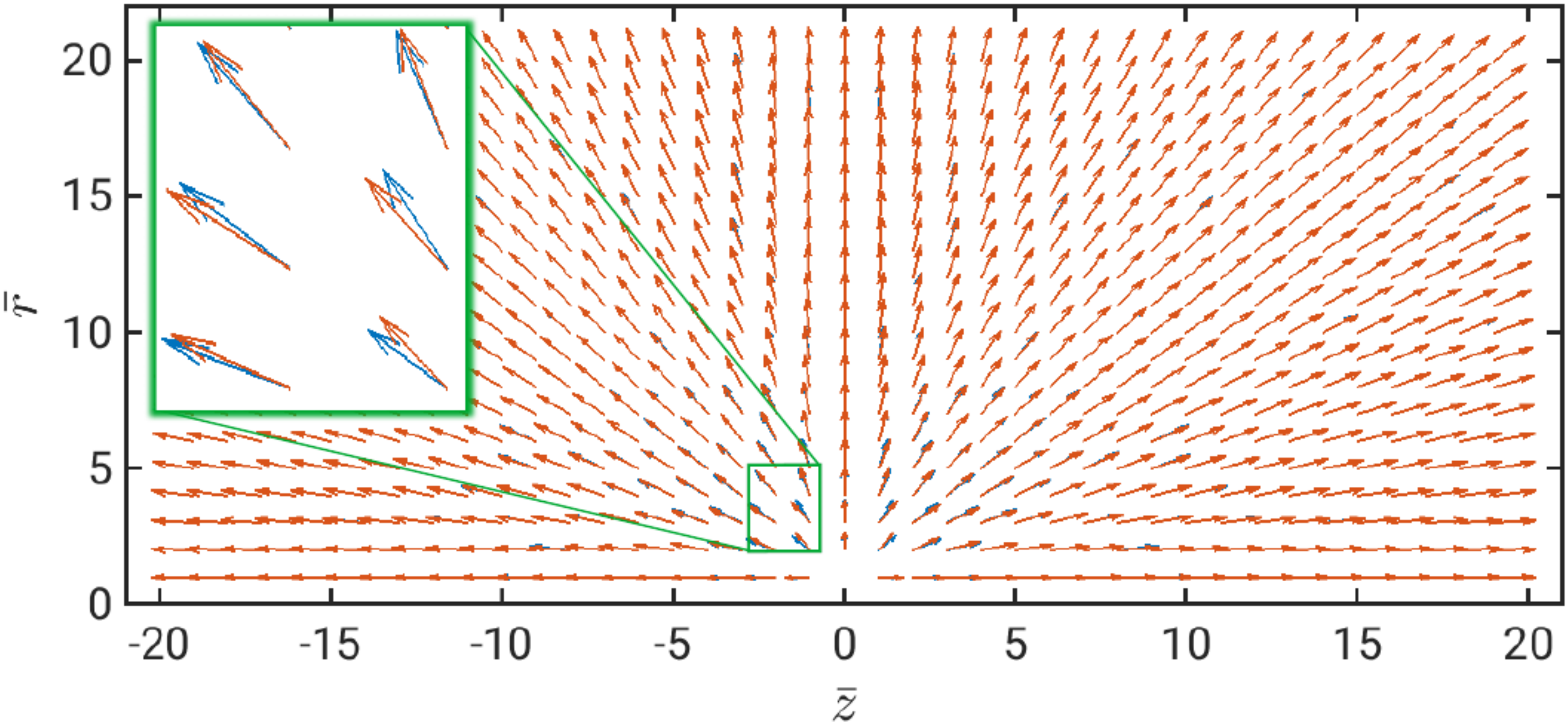}
\caption{Shown are the vector components $\vec q=(q_r, q_z)$, to allow for a comparison of the monopole configurations before (exact, blue) and after (with error, brown) the minimisation procedure. The magnification on the left illustrates the differences more clearly.}
\label{fig:VectorWaves}
\end{figure}

\begin{figure}[H]
\centering
\includegraphics[width=0.8\textwidth]{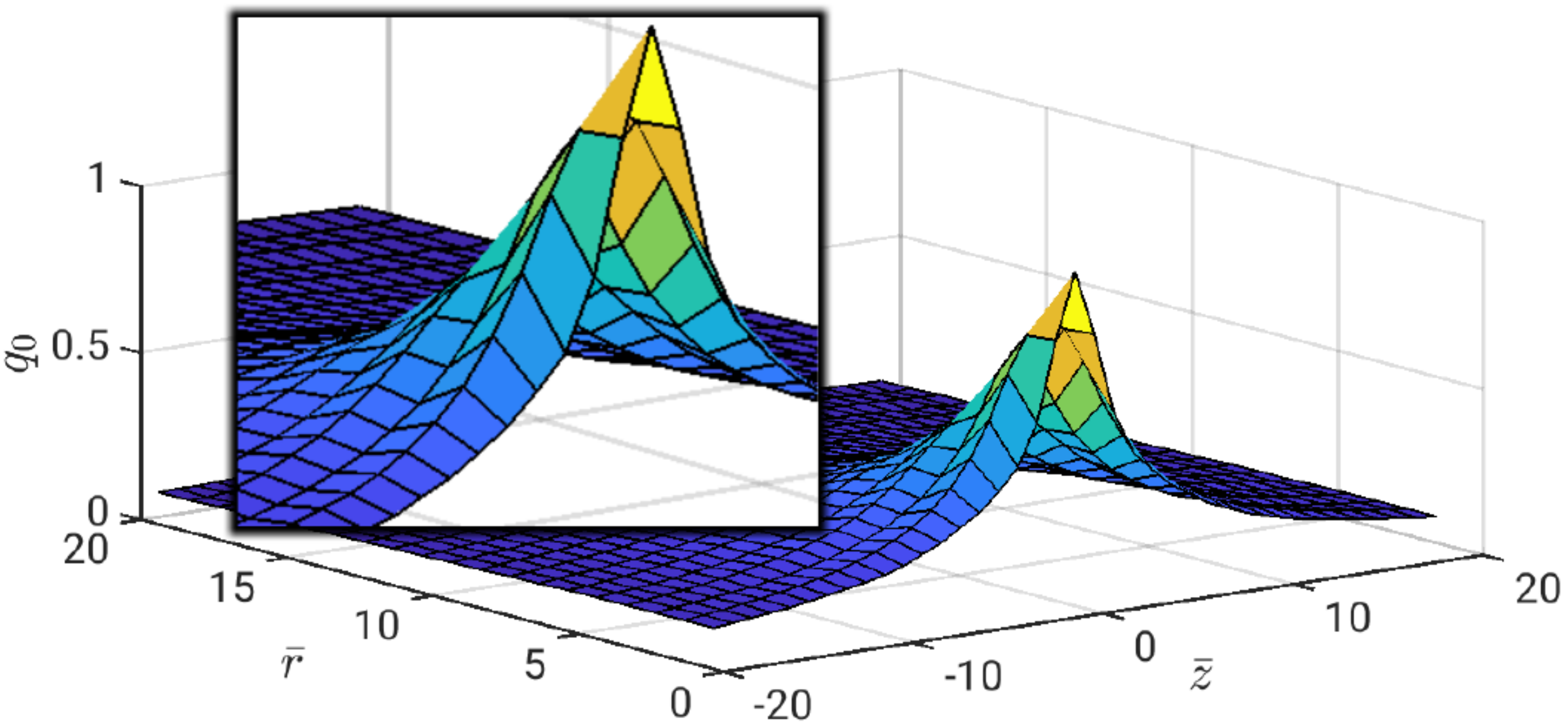}
\caption{$q_0$-field in the $rz$-plane for the minimised monopole configuration. Observe the wave-like discontinuities along the radial direction.}
\label{fig:ScalarWaves}
\end{figure}

In Fig.~\ref{fig:VectorWaves} we compare the two-component vectors $(q_r,q_z)$ in the box before and after the minimisation and find clear differences between the numerical and analytical minima. Further, it can be clearly seen in Fig.~\ref{fig:ScalarWaves}, that after the minimisation the $q_0$ component showed up wave-like discontinuities. These deviations from the analytical solution are not physical and need to be avoided. From the analytical solution, we know that the field should be smooth.

The contribution of the discontinuities to the overall energy is very tiny and for bigger lattices they get negligible small. Nevertheless they are always present in the computations and are dangerous for the determination of static dipole configurations. Even fixing the monopole centres these small fluctuations lead to tedious collapses of the monopoles as was shown in several diploma theses \cite{Resch,Theuerkauf,Anmasser}.

\subsection{Solution to the numerical instability}\label{SolutionToTheNumericalInstability}

As shown in Fig.~\ref{fig:VectorWaves}, the problem discussed in the previous section is accompanied by misalignments of the $\vec q$-vectors. This fact suggests to diminish the variations in the directions of the $\vec q$-field. Therefore, we have added a third term in Eq.~(\ref{eqn:massTerm}) to the energy functional on the lattice in order to minimise the fluctuations of $\vec n:=\vec q/|\vec q|$ and forcing discontinuities to vanish in the minimisation procedure. The energy functional which we minimise is therefore
\begin{equation}\label{eqn:massTerm}
\bar H_\mathrm{mod}^\lambda:=\bar H_\mathrm{tot}+\underbrace{\lambda\sum_{\bar i}
\left(\partial_{\bar i}\frac{\vec q}{|\vec q|}\right)^2}_{\bar H^\lambda}\,, \quad
\bar i=\{\bar r,\bar z\},
\end{equation}
where the order of magnitude is controlled by $\lambda$. This extra term $\bar H^\lambda$ reminds us of the mass term for the pion in the Skyrme model. Likewise, the physical interpretation of our new term is, that it gives mass to photons and hence suppresses photonic excitations. Since we are computing static snapshots, the application of this method is justified. We are not interested in the size of $\bar H^\lambda$, we want to use it as a constraint and to minimise among the set of lattice configurations with minimal $\bar H^\lambda$.

\begin{figure}[H]
\centering
\includegraphics[width=0.8\textwidth]{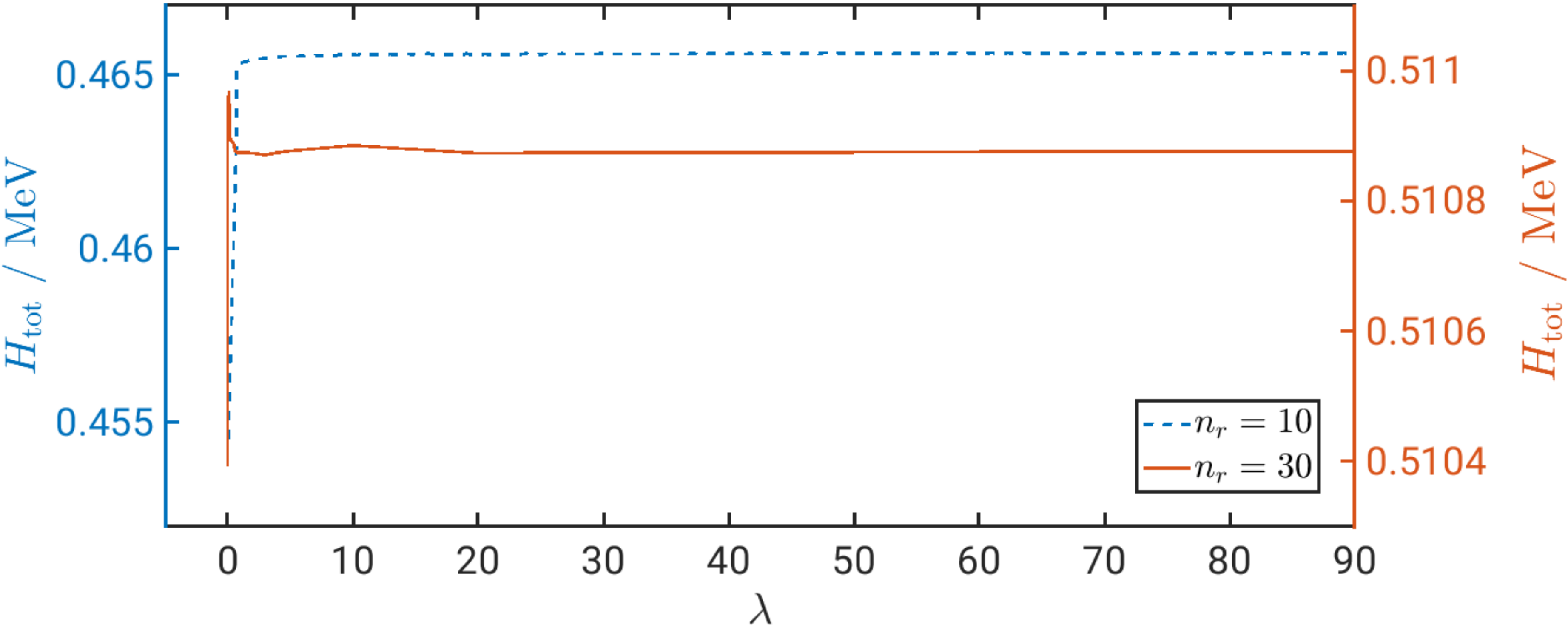}
\caption{Plot of the total energy $H_\mathrm{tot}$ versus $\lambda$ for lattices of size $n_r\times2n_z$ with $n_r=n_z=10$ and 30. Increasing $\lambda$, discontinuities get smoothed out in both cases very soon. By choosing a high enough value, they vanish completely and rising it further has no effect. The difference in the left and right scales is caused by the much smaller accuracy for $n_r=10$.}
\label{fig:EnergyVsLambda}
\end{figure}

\begin{figure}[H]
\centering
\includegraphics[width=0.8\textwidth]{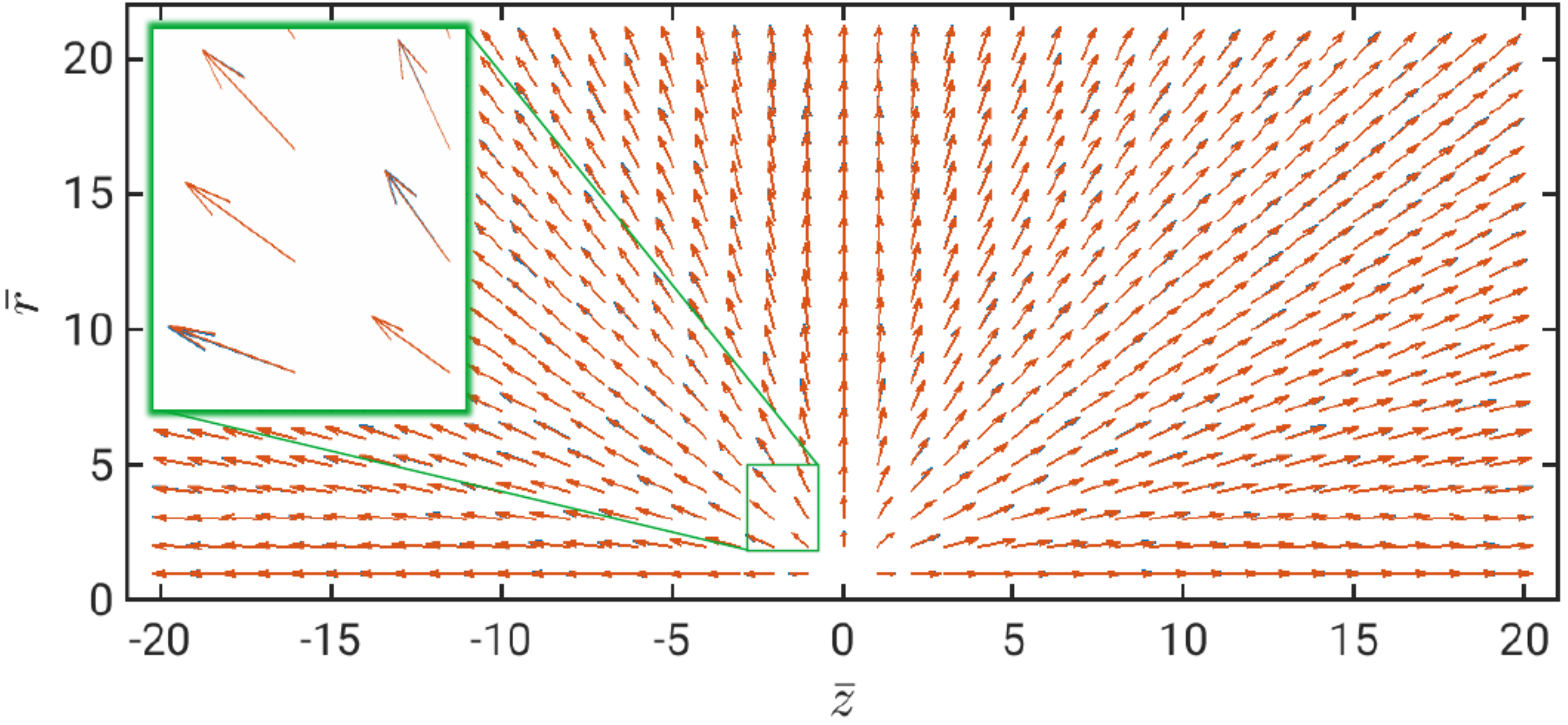}
\caption{Same simulation and parameters as in Fig.~\ref{fig:VectorWaves}, but using in the minimization procedure the additional term $\bar H^\lambda$ of Eq.~(\ref{eqn:massTerm}) at $\lambda = 100$. As can be seen, the vector part of the configuration is nearly indistinguishable before and after the minimisation.}
\label{VektNachKorr}
\end{figure}

\begin{figure}[H]
\centering
\includegraphics[width=0.8\textwidth]{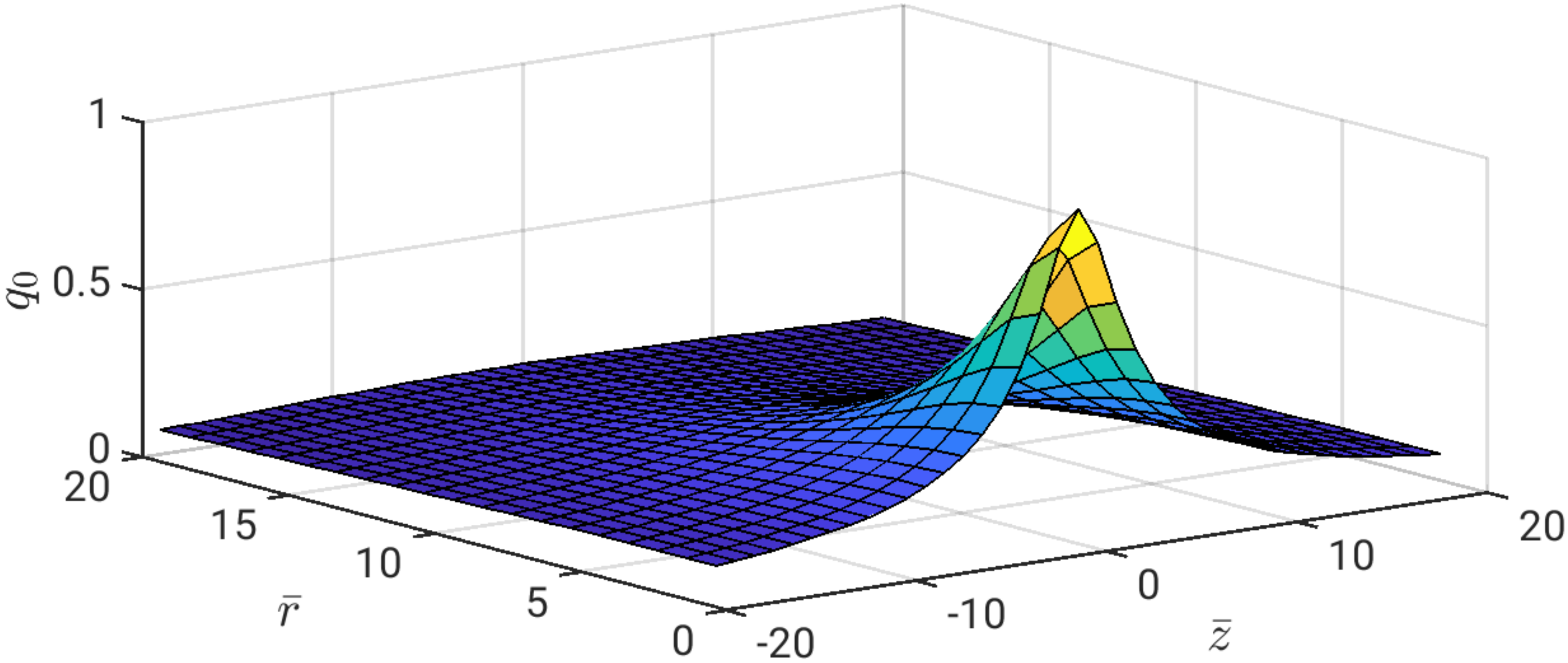}
\caption{Same simulation and parameters as in Fig.~\ref{fig:ScalarWaves}, but with the additional term $\bar H^\lambda$ at $\lambda = 100$. Observe, that the waves in radial direction have vanished.}
\label{fig:ScalarWavesSmooth}
\end{figure}

In Fig.~\ref{fig:EnergyVsLambda}, we show a plot of $H_\mathrm{tot}$ for simulations with different values for $\lambda$ for $10 \times2\cdot10$ and $30 \times2\cdot30$ lattices. As one may expect, we see that the results do not change significantly any more, if the parameter $\lambda$ exceeds a certain value. With this method the two-component vectors before and after the minimisation show no visible difference as Fig.~\ref{VektNachKorr} demonstrates at $\lambda=100$. The correction of the minimisation procedure by the additional term leads to a nicely smooth distribution of $q_0$-values around the centre of the monopole, see Fig.~\ref{fig:ScalarWavesSmooth}.

\begin{figure}[H]
\centering
\includegraphics[width=0.8\textwidth]{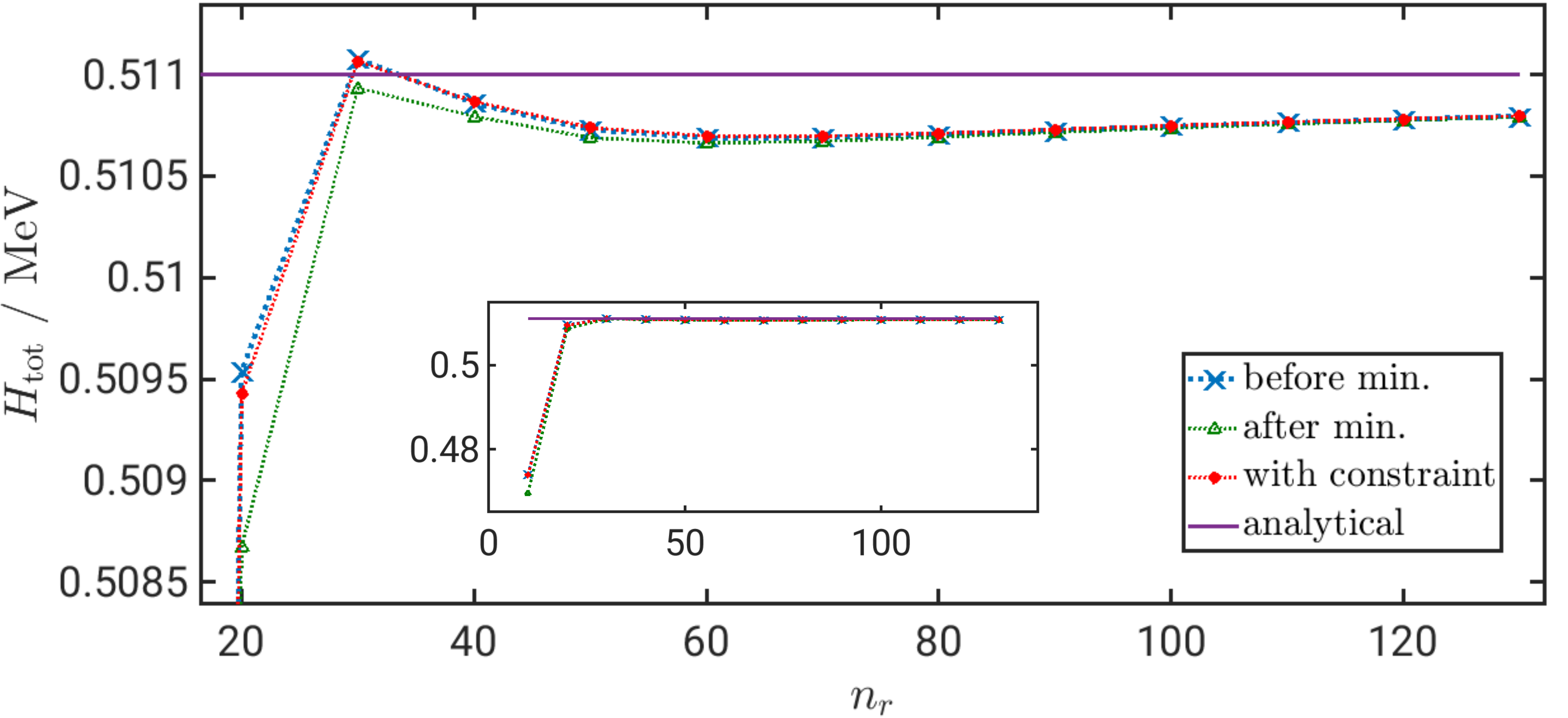}
\caption{Total energy $H_\mathrm{tot}$ of a monopole configuration with $Z=10\,r_0$ and $R=10\,r_0$ and different lattice sizes $n_r\times2n_z$ and $n_r=n_z$, determined with different methods as explained in the text. The inset shows that the energy scale was chosen in order to show the tiny differences in the results clearly.}
\label{fig:EtotFinish}
\end{figure}

In Fig.~\ref{fig:EtotFinish} we compare the results of the different numerical evaluations of the total energy of the monopole for $Z=10\,r_0$ and $R=10\,r_0$ and various lattice sizes $n_r\times2n_z$ and $n_r=n_z$. We can clearly see that deviations from the analytical results due to the lattice discretization. As discussed in Sect.~\ref{PrecisionOfTheConjugateGradientMinimization}, the minimisation without constraint leads to a strange shape of the field distribution but to a small effect on the energy only, as can be seen comparing the results before and after minimisation. With the additional constraint~(\ref{eqn:massTerm}) the minimal configurations shows the expected behaviour and the expected energy. The results nicely agree before and after the minimisation with the constraint. Remaining differences to the energy of the analytical configuration are due to the chosen discretization of the energy functional and the missing terms for the potential and tangential energy outside the box. Adding the analytical expressions  $2H_\mathrm{pot}^\mathrm{out}\ist{schaetzwertPotOut}0.312~\mathrm{keV}$ for the sum of both terms would lead to a small overshooting of the analytical result.

\begin{figure}[H]
\centering
\includegraphics[width=0.8\textwidth]{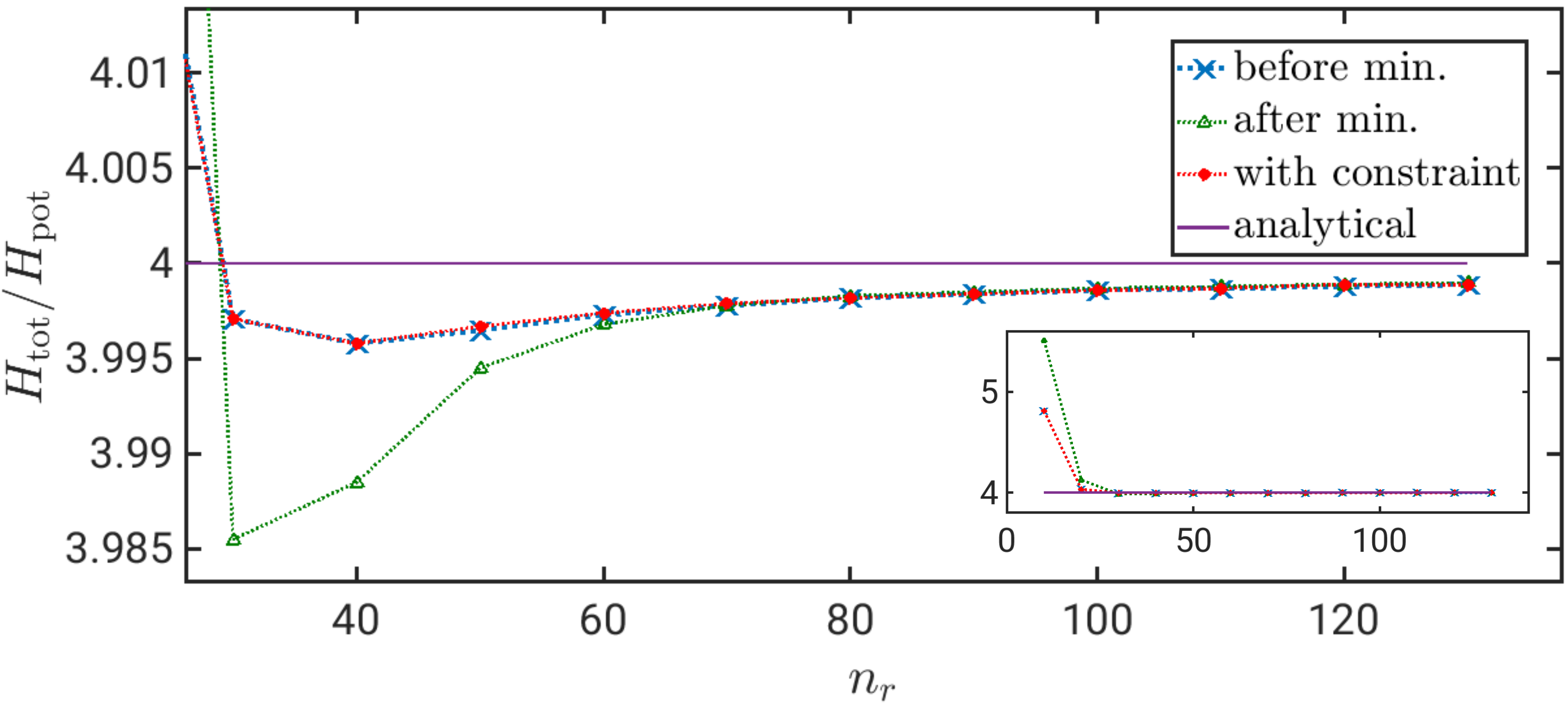}
\caption{Same situations as Fig.~\ref{fig:EtotFinish}. The ratio $H_\mathrm{tot}/H_\mathrm{pot}$ is shown, which is $4$ for the analytical minimum configuration.}
\label{fig:RatioFinish}
\end{figure}

A similar comparison we show for the ratio $H_\mathrm{tot}/H_\mathrm{pot}$ in Fig.~\ref{fig:RatioFinish}. Due to the Hobart-Derrick theorem this value should approach $4$.

\section{Conclusions}\label{Sec:Schluss}
Only the three field variables of a scalar SO(3)-field and an appropriate Lagrangian with two terms are necessary to describe the behaviour of topological solitons with long-range interaction. SU(2), the double covering group of SO(3), is well suited for investigations of the topology of field configurations. The known solutions of the field equations gives a nice opportunity to investigate the problems appearing in the numerical computations. Several attempts~\cite{Resch,Theuerkauf,Anmasser} to determine field distributions of static charges in two~\cite{Theuerkauf,Anmasser} and three~\cite{Resch} dimensions had failed due to numerical instabilities. We have shown in this article that the origin of these instabilities are misalignments of the rotational axes corresponding to the scalar SU(2)-field. Suppressing these wave-like disturbances with a corresponding constraint we were able to get results of the numerical minimisation procedure in excellent agreement with the analytical solutions. The application of this method to two soliton systems will in future be of special importance. It will allow to get accurate results for the interaction at short distances. Deviations from the Coulomb behaviour are expected and should be compared with the the running coupling, well-known from field theory.
%\section{Acknowledgements}

\bibliography{literatur}
\bibliographystyle{unsrt}
\end{document}